\documentclass[preprint,groupedaddress]{revtex4}

\usepackage{graphicx}
\usepackage{dcolumn}
\usepackage{bm}
\usepackage{color}

\usepackage{amssymb}
\usepackage{amsthm}
\usepackage{amsmath}
\usepackage{amsfonts}
\usepackage{pdfpages}
\usepackage{epstopdf}

\begin{document}


\title{Impact of direct digital synthesizer finite resolution on atom gravimeters}

\author{R. Karcher}
\author{F. Pereira Dos Santos}
\author{S. Merlet}

\email[]{sebastien.merlet@obspm.fr}
\affiliation{LNE-SYRTE, Observatoire de Paris, Universit\'e PSL, CNRS, Sorbonne Universit\'e,
61 avenue de l'Observatoire, F-75014 Paris, France}

\date{\today}

\begin{abstract}

We report on the study of the impact of the finite resolution of the chirp rate applied on the frequency difference between the Raman lasers beamsplitters onto the phase of a free fall atom gravimeter. This chirp induces a phase shift that compensates the one due to gravity acceleration, allowing for its precise determination in terms of frequencies. In practice, it is most often generated by a direct digital synthesizer (DDS). Besides the effect of eventual truncation errors, we evaluate here the bias on the $g$ measurement due to the finite time and frequency resolution of the chirp generated by the DDS, and show that it can compromise the measurement accuracy. However, this effect can be mitigated by an adequate choice of the DDS chirp parameters resulting from a trade-off between interferometer phase resolution and induced bias.     

\end{abstract}


\maketitle

\section{Introduction}\label{intro}

Light pulse atom interferometry is now a mature technique, specially fit to design inertial sensors \cite{Riehle1991, Kasevich1991} which compete with state of the art classical sensors \cite{Gillot2014, Freier2016, Asenbaum2017, Savoie2018}. Among these, Cold Atom Gravimeters (CAG) based on three pulse Mach-Zehnder sequence \cite{Kasevich1991} are absolute gravimeters which have demonstrated the ability to perform continuous measurements with better short term sensitivities \cite{Gillot2014,Peters1999, Hu2013} and long term stabilities than their classical counterparts \cite{Freier2016}.  They have also reached comparable or better accuracies by the careful evaluation of their systematic effects. The control of these systematic effects crucially depends on the control of the phase difference between the light beamsplitters, which get imprinted onto the atomic wavepackets at each pulse during the interferometer. Both the spatial and temporal laser phase fluctuations, which have been extensively studied in the past, such as in \cite{Peters2001, Louchet2011, Schkolnik2015, Tao2015, Zhou2016, Karcher2018}, need to be perfectly controlled to ensure the accuracy of the measurement of the local gravity acceleration $g$.

In gravimeters based on atom interferometry, the measurement of $g$ is derived from the determination of the Doppler frequency chirp induced by the free fall of the atoms onto the lasers \cite{Peters2001}. In practice, lasers are kept on resonance by sweeping over the interferometer duration their frequency difference thanks to an agile and stable oscillator. Any parasitic frequency or phase shifts with respect to the ideal chirp will thus induce errors on the $g$ measurement, such as for instance the bias due to frequency dependent radio-frequency delays in the electronics studied in \cite{Peters2001}, which can be rejected by a proper symmetrization of the chirp with respect to the mid pulse between both directions of the effective wave vector. 
In the following, we investigate the impact of the finite frequency resolution of the oscillator used to chirp the lasers frequency difference during the interferometer. This chirp being synthesized with discrete steps out of a clock reference signal which we consider here as perfectly stable, produces periodic frequency errors, leading to periodic errors on the interferometer phase. After a brief description of our measurement method of $g$, we first calculate the biases in the $g$ measurements induced by truncation and frequency errors arising from the finite resolution of the direct digital synthesizer (DDS) we use. Then, we performed measurements of these biases, which we find in agreement with numerical calculations. This analysis allows us to accurately evaluate the impact of these effects on our absolute cold atom gravimeter.

\section{Numerical simulations: frequency error of the chirp}

\subsection{$g$ measurement protocol}

In the most mature atom gravimeters, $g$ measurements are performed via Raman interferometry on free falling atoms. There, the interferometer is realized thanks to a sequence of three two-photon stimulated Raman transitions of duration $\tau-2\tau-\tau$ separated by a free evolution time $T$, which respectively split, redirect and finally recombine the atomic wave packets. As mentioned above, during the free fall, the Doppler shift modifies the resonance condition which has to be compensated for by applying a frequency chirp to the frequency difference between the Raman lasers. This chirp rate $\alpha$ adds a phase shift ($\alpha T^2$) to the interferometer that, when properly tuned ($\alpha=\alpha_0=k.g$), exactly compensates the phase shift induced by the gravity acceleration ($-k.gT^2$), $k$ being the effective wave vector of the Raman transition. Remarkably, this leads to a dark fringe in the fringe pattern obtained when scanning $\alpha$, whose position does not depend on the interferometer duration \cite{Farah2014}. The measurement of the gravity acceleration, and its fluctuations, can thus finally be obtained from the precise tracking of this dark fringe and from the determination of the corresponding value of the chirp rate $\alpha_0$. Many of the systematic effects, such as related to one photon light shifts or the quadratic Zeeman effect, are rejected by averaging $g$ measurements over opposite directions of the effective wave vector $k$ \cite{Louchet2011,Weiss1994}. Indeed, unlike the gravity phase shift, these systematics do not depend on the direction of $k$ \cite{Peters2001, Louchet2011}. However, the efficiency of this rejection depends on the superposition of the two trajectories for the two interferometer configurations. In practice, the difference in the momenta imparted to the atoms leads to small differences between the trajectories, of about a few mm. These are much smaller than the few tens of centimetres of the free fall distance, which guarantees the efficiency of this rejection technique.
 
\subsection{Truncation error}

In our experiment, we generate the chirp thanks to a direct digital synthesizer (AD9852, from Analog Devices). It is a 48-bits DDS, clocked at $300$MHz, which corresponds to a frequency resolution of about $300$MHz/2$^{48} \simeq 1\mu$Hz. Within its chirped mode of operation, we can control both the temporal step $\delta t$ and the frequency step $\delta \nu$, which can be tuned to design the desired chirp with a rate $\alpha$ (inserted picture in figure \ref{fig1}). In our case, the frequency chirp required to compensate for gravity, of about 25~MHz/s, is realized with frequency steps of $125~$Hz every $10~\mu$s. Indeed, the DDS is compared in a phase frequency detector to the beat note between the Raman lasers (after a down conversion, realized by mixing the beat note with a $7~$GHz oscillator, and a subsequent division by a factor $2$). At last, this results in a frequency chirp resolution of about $2 \times 1\mu Hz/10\mu s \simeq 0.2~$ Hz/s. The corresponding resolution in terms of gravity acceleration is $8.3\mu$Gal ($1~\mu$Gal$=10~$nm/s$^{2}$), which is smaller than the shot to shot noise on the $g$ measurement (peak to peak of about $\sim 100\mu$Gal). This resolution is thus not a limit in our measurement protocol, which consists in a digital integrator that steers the chirp rate onto the central fringe \cite{Merlet2009}. However, the correction applied by the lock system onto the chirp rate is impacted by this resolution, as there is a difference between the corrections we calculate at each measurement cycle, and the ones that are actually applied, due to truncation errors. These errors, which amount on average to $1/2$ bit (or $\sim 4.15~\mu$Gal), could in principle be eliminated by recording the applied changes instead of the requested ones. On the other hand, and quite remarkably, these errors cancel when averaging over the two opposite $k$ directions, as they lead to underestimating the chirp rate when the frequency is ramped up and overestimating it when ramped down.
 
Finally, more important than the effect of these truncations, the DDS cannot exactly produce the required change of the frequency$/$phase of the Raman phase difference because of its step wise character. This leads to systematic errors, which we discuss and evaluate in the following section.
 
\subsection{Frequency error: finite resolution of the DDS}\label{Calcul}

\begin{figure}[h!]
 \centering
 \includegraphics[width=9cm]{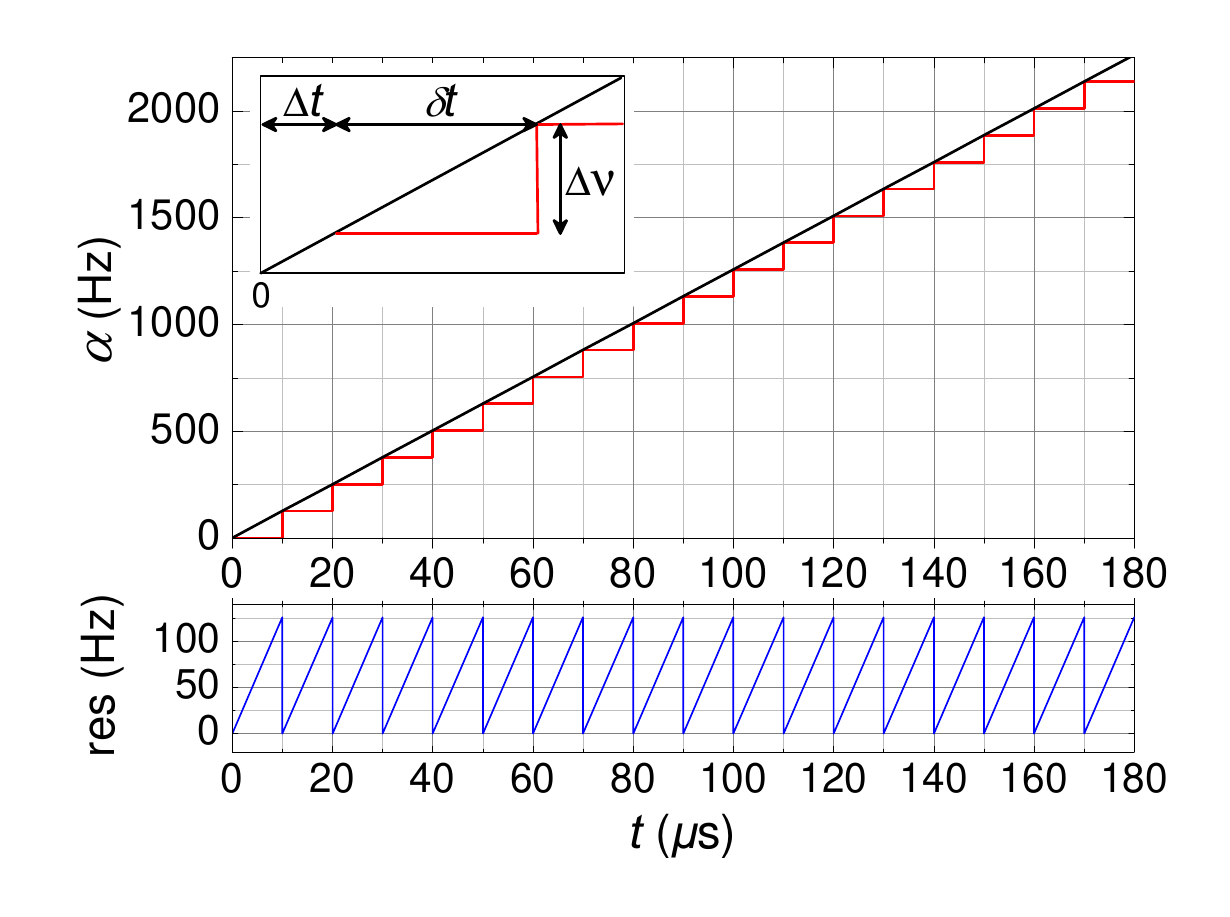}
 \caption{Frequency chirp applied to the Raman lasers frequency difference. The perfect linear case is represented in black and the actual chirp, composed of frequency steps of $\delta\nu=125$~Hz and temporal steps $\delta t=10~\mu$s, in red. The insert shows the parameters of the DDS ($\delta t$, $\delta \nu$), $\Delta t$ being the adjustable delay between the starts of the interferometer and the chirp.  Bottom: residual or frequency difference $\Delta f$ between the ideal and actual chirps, represented as a sawtooth function.}
   \label{fig1}
\end{figure}

As shown in figure \ref{fig1}, the chirp is realized by the DDS by incrementing the laser frequency by $\delta \nu$ every $\delta t$, leading to a chirp rate $\alpha$. The deviation from a perfect linear frequency chirp induces a periodic frequency error represented by the sawtooth function displayed on figure \ref{fig1}. This function is determined by the two parameters $\delta \nu$ and $\delta t$. Also, we note $\Delta t$ the time difference between the beginning of the chirp and the beginning of the interferometer sequence.

To calculate the bias on the interferometer phase, we use the sensitivity function $g_s$ \cite{Cheinet08}, which describes the impact of frequency fluctuations $\Delta f(t)$ onto the interferometer phase $\Delta\Phi$:

\begin{equation}\label{g_expression}
\Delta \Phi = \int_{-\infty}^{+\infty} g_s(t) 2\pi \Delta f (t) dt
\end{equation}

It is an odd function given, for $t>0$, by :

\begin{equation}
g_\text{s}(t) = \left\{\begin{array}{ll}
            \sin (\Omega_R t) & \textrm{for $0<t<\tau$}\\
            1 & \textrm{for $\tau<t<T+\tau$}\\
            -\sin (\Omega_R (T - t)) & \textrm{for $T+ \tau<t<T+2\tau$}\\
            0 & \textrm{for $t>T+2\tau$}\\
            \end{array} \right.
\end{equation}

From now on and throughout the rest of the paper, we take, without loss of generality, the time origin $t=0$ at the centre of the middle $\pi$ pulse. $\Omega_R$ is the Rabi frequency, given by $\Omega_R = \pi/2\tau$.

To evaluate the bias on the $g$ measurement, we thus simply integrate equation \eqref{g_expression} modulated by the sawtooth function over the duration of the interferometer. 

\begin{figure}[h!]
 \centering
 \includegraphics[width=8.5cm]{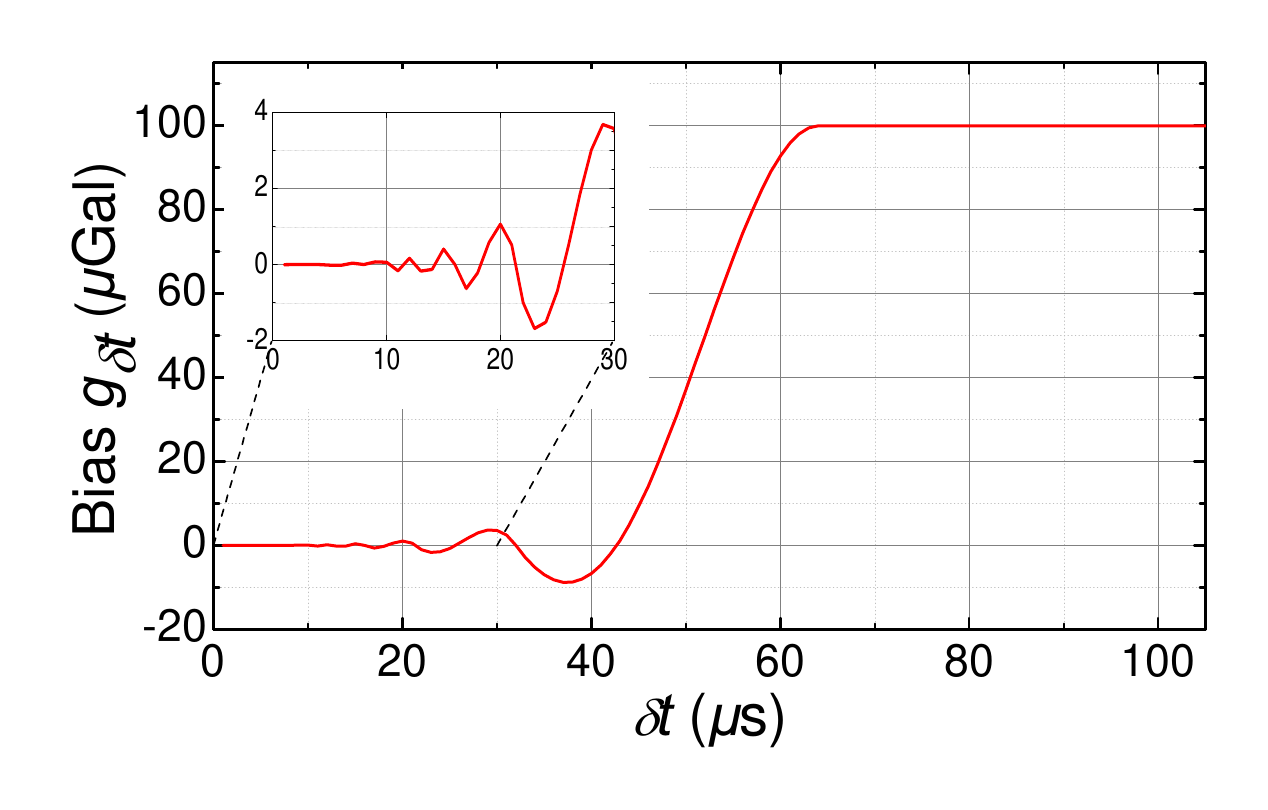}
 \caption{Calculated bias $g_{\delta t}$ as a function of the temporal step of the chirp $\delta t$ for $T=80~$ms and $\tau=16~\mu$s. Focus on the $[0-30]~\mu$s interval.}
 \label{fig3}
\end{figure}

Figure \ref{fig3} shows the calculated bias for $\delta t$ ranging from $1\mu$s to $100~\mu$s for our typical interferometer parameters ($\tau = 16\mu~$s, $T= 80~$ms). For $\delta t = 10\mu$s the expected bias ($0.06~\mu$Gal) is found to be negligible, but it increases as a function of $\delta t$, leading to a bias as large as $\sim 100~\mu$Gal for $\delta t =100~\mu$s.

For a given $\delta t$, two parameters can be easily modified to modulate the impact of frequency errors :

\begin{itemize}

\item First, one can vary the delay $\Delta t$ between the beginning of the interferometer with respect to the beginning of the chirp. This parameter is varied between 0 and $100~\mu$s on figure \ref{SimuOffset}, which displays the results of the calculated bias as a function of $\Delta t$ for $\delta t= 10~\mu s\, ,20~\mu s \, ,50~\mu s \, \text{and} \, 100~\mu s $.

\begin{figure}[h!]
   \begin{minipage}[c]{.46\linewidth}
      \includegraphics[width=8.5 cm]{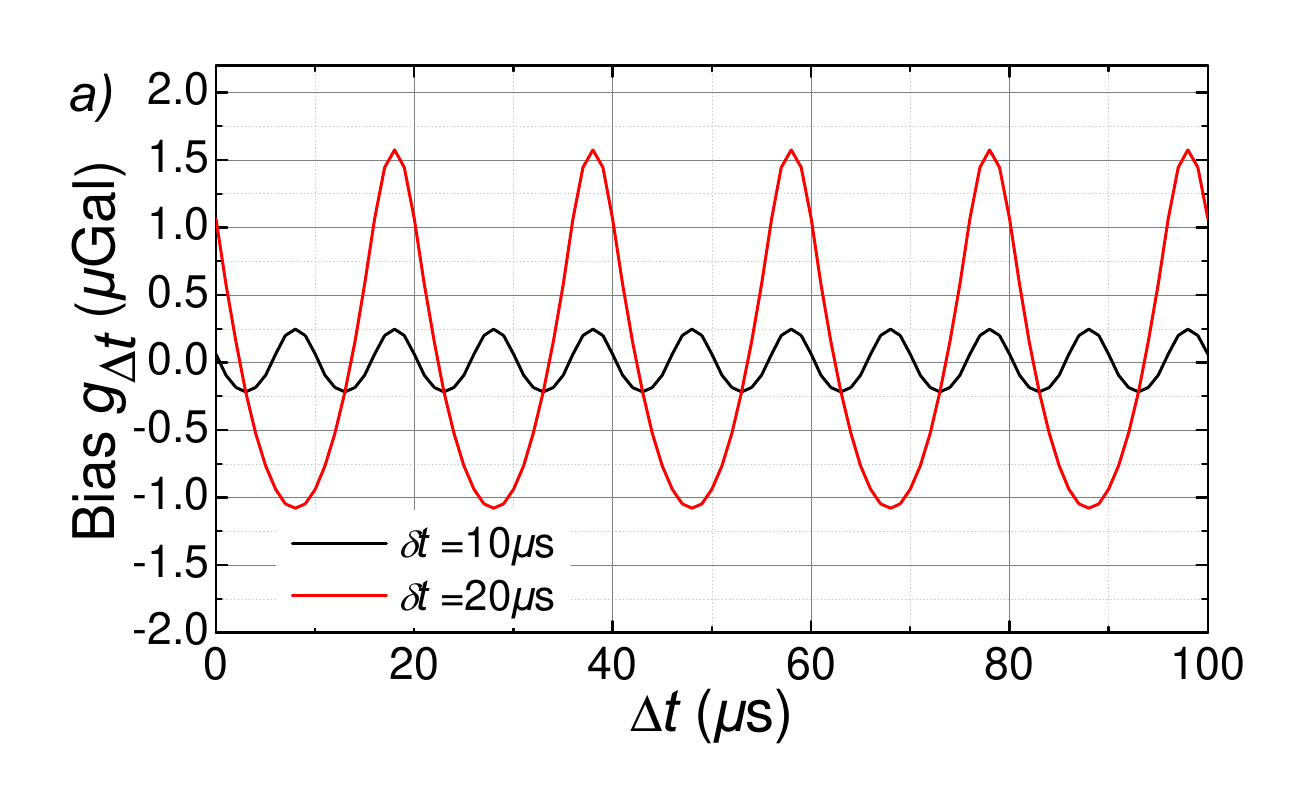}
   \end{minipage} \hfill
   \begin{minipage}[c]{.46\linewidth}
      \includegraphics[width=8.5 cm]{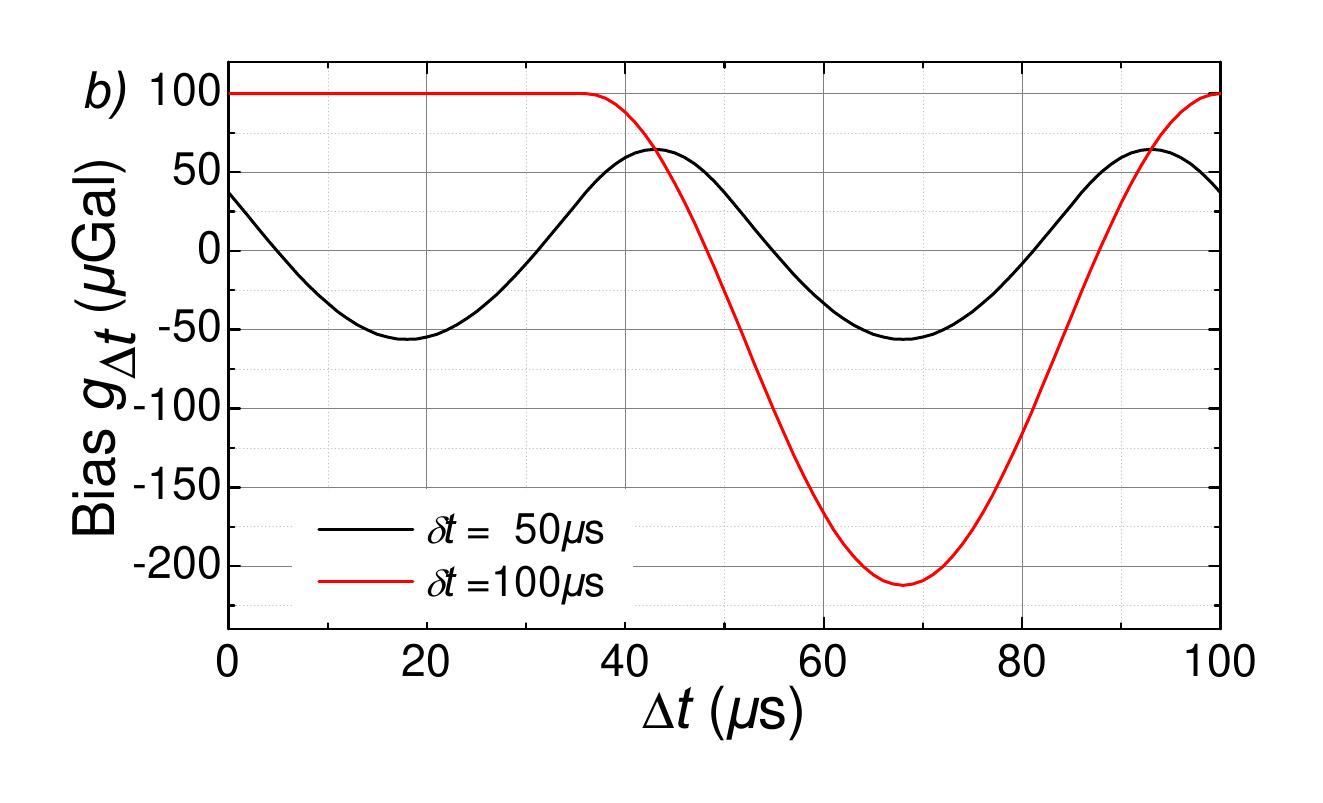}
   \end{minipage}
   \caption{Calculated bias $g_{\Delta t}$ as a function of the delay $\Delta t$ between the chirp sequence and the interferometer sequence for  a) $\delta t= 10~\mu$s in black and $20~\mu$s in red. b) $\delta t= 50~\mu$s in black and $100~\mu$s in red.}
   \label{SimuOffset}
\end{figure}

As expected, the results are periodic with a period $\delta t$. The amplitude of the bias is lower than $2~\mu$Gal for $\delta t \leq 20~\mu$s but reaches over $-200~\mu$Gal for $\delta t = 100~\mu$s. Note that one can null the bias for specific values of the delay $\Delta t$, which depend on $\delta t$.

\item  Second, one can change the Raman pulse durations. Figure \ref{SimuTau} is the result of the calculated bias as a function of $\tau$ for $\delta t = 50~\mu$s and 100~$\mu$s. To illustrate the effect, we choose $\delta t$ for which the effect is large. In this calculation, we use realistic durations of $\tau$ of the order of few tenth of $\mu$s and there is no delay ($\Delta t =0$). The result is a periodic function which dampens for increasing values of $\tau$. This illustrates the fact that when $\tau \gg \delta t$, the bias averages down to zero.

\begin{figure}[h!]
   \begin{minipage}[c]{.46\linewidth}
      \includegraphics[width=8.5 cm]{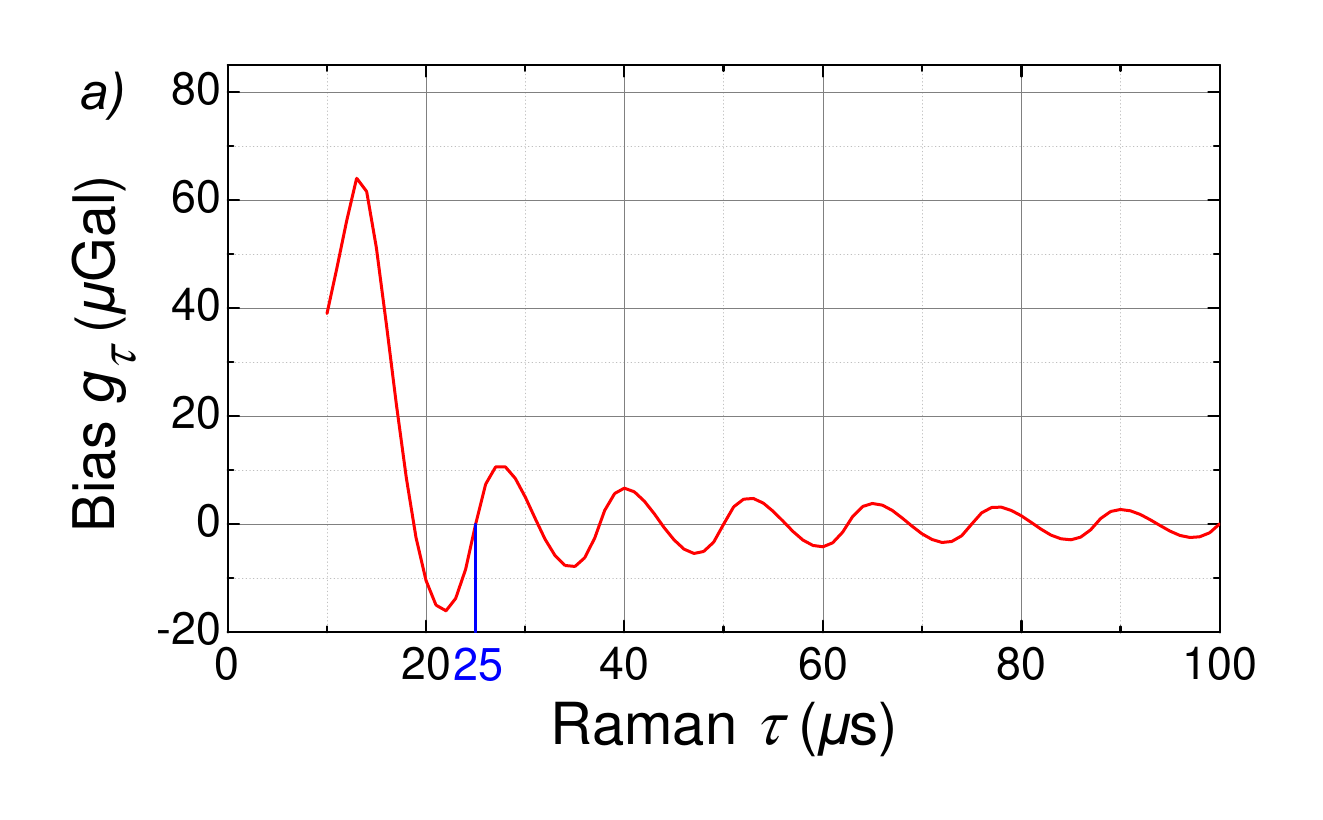}
   \end{minipage} \hfill
   \begin{minipage}[c]{.46\linewidth}
      \includegraphics[width=8.5 cm]{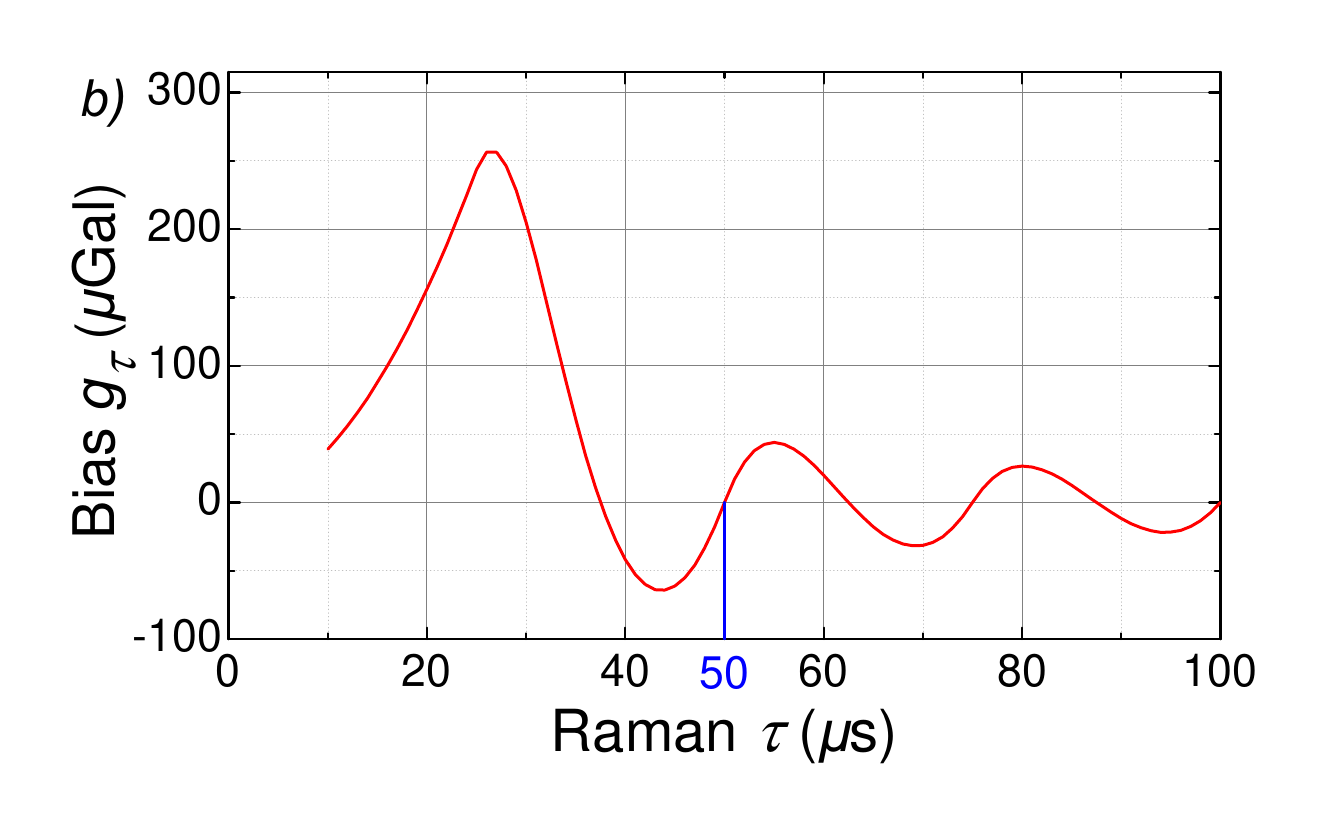}
   \end{minipage}
   \caption{Calculated bias $g_{\tau}$ as a function of the Raman pulse duration $\tau$ for a) $\delta t =~50\mu$s and b) $\delta t=~100\mu$s. The case where the duration $\tau$ equals half $\delta t$, for which the bias is null, is represented in blue.}
   \label{SimuTau}
\end{figure}

In a different manner than the truncation error which is rejected by the $k$-reversal algorithm, this bias can be cancelled with a proper choice of the Raman pulse duration. A simple choice would be to use $\tau = \delta t/2$, for which this effect is calculated to be null.

\end{itemize}

\section{Measurements and comparison with numerical calculations.}

In this section, we report on the experimental study of the impact of the DDS finite resolution on the measurements of our cold atom gravimeter, which we compare to the calculations described in the previous section.

\subsection{Experimental setup}

We present here briefly the experimental setup we have used to perform the measurements. A more detailed description can be found in \cite{Louchet2011}. We start by trapping a sample of $^{87}Rb$ atoms in a 3D MOT, and further cool it down to $2~\mu$K with far detuned molasses. About $10^8$ atoms are released in free fall for 200~ms and prepared in the $\vert m_F=0\rangle$ magnetic state in a narrow vertical velocity distribution. We then apply the interferometer sequence which lasts a total time of $2T=160$~ms. As discussed earlier, the Raman lasers are kept resonant with the atoms during the free fall, by actually chirping one of the two Raman lasers \cite{Cheng2015}, so as to compensate for the linearly increasing Doppler shift.

The phase difference in this two-wave interferometer modulates the population $N_1$ and $N_2$ at the two output ports, which are measured at the bottom of the chamber by a internal-state selective fluorescence detection, using the state labelling property of Raman transitions \cite{Borde89}. From these measurements we derive the transition probability $P$, which is given by $P=N_1/(N_1+N_2)=\frac{1}{2}\left[ 1+C \cos(\Delta\Phi)\right] $, where $C$ is the contrast of the interferometer and $\Delta \Phi= -kgT^2+\alpha T^2$ the total interferometer phase shift. Gravity measurements are performed by steering $\alpha$ towards $kg$, so as to nullify $\Delta \Phi$. The measurement repetition rate is about 3Hz.
 
In the following, the measurements are performed in a differential way by alternating gravity measurements with different sets of parameters $S = \lbrace \delta t, \Delta t , \tau \rbrace$ and directions of the effective wave vector. Our conventional parameters are $S_{ref}=\lbrace 10\mu s, 0\mu s, 16\mu s \rbrace$.

In practice, we compare a first pair of configurations, $g(k_{\uparrow},  S_{ref} )$ and $g(k_{\downarrow} ,S_{ref})$, with opposite effective wave vectors and reference sets of parameters, with other pairs of configurations $g(k_{\uparrow},  S )$ and $g(k_{\downarrow} ,S)$. In most of the measurements presented below, the quantity of interest is the difference between the average measurements of the pairs:

\begin{equation}
\label{eqDeltag}
\Delta g = \frac{g(k_{\uparrow}, S_{ref}) +g(k_{\downarrow}, S_{ref})}{2} - \frac{g(k_{\uparrow}, S) +g(k_{\downarrow}, S)}{2}
\end{equation}

\subsection{Evaluation of the truncation error}

At first, we tried to evaluate the truncation error. As it is rejected by the average over opposite directions of $k$, we consider here the differences between single-$k$ directions such as:

\begin{equation}
\Delta g_{TE} = \frac{\left[ g(k_{\uparrow}, S_{ref})-g(k_{\uparrow}, S)\right]-\left[g(k_{\downarrow}, S_{ref}))-g(k_{\downarrow}, S)\right]}{2}
\end{equation}

\begin{figure}[h!]
 \centering
 \includegraphics[width=9cm]{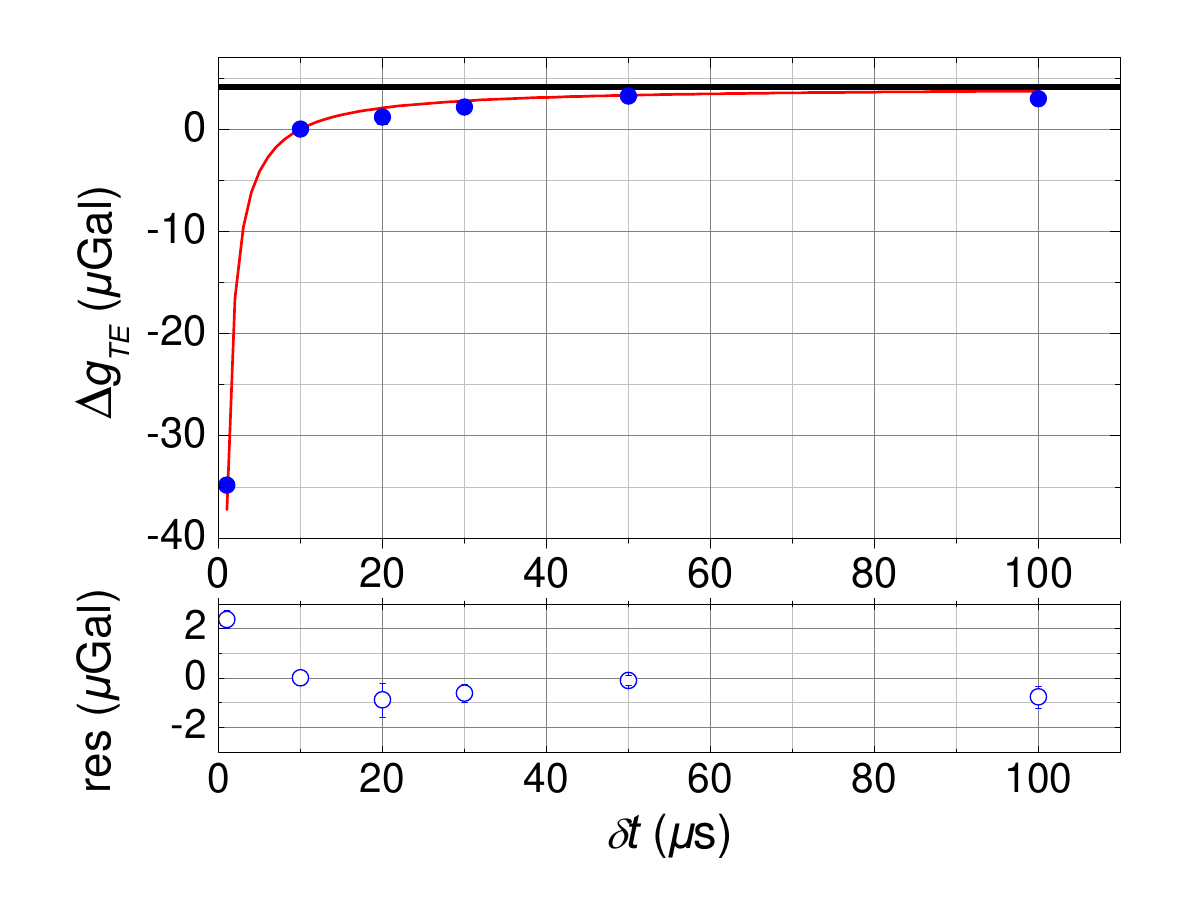}
 \caption{Measurements of the truncation error $\Delta g_{TE}$ (in blue) as a function of the temporal step of the DDS $\delta t$. The results of the calculation are displayed as a continuous red line. The asymptotic line, which corresponds to a null bias and to a difference of $4.15~\mu$Gal with respect to the reference configuration of $\Delta t = 0~\mu$s, $\tau=16~\mu$s, is represented in black. The residues between theory and measurements are displayed at the bottom as blue open circles. }
 \label{Troncerror}
\end{figure}

Results are represented on figure \ref{Troncerror} (taking as a reference $\Delta g_{TE}^{\delta t = 10 \mu s}$ for which the bias is expected to amount to $\approx 4.15~\mu$Gal), together with the calculated errors in red.

These differential measurements are found to converge towards $4.15~\mu$Gal, corresponding in fact to a null bias for asymptotically large values of $\delta t$. The measurements agree with the calculated values, within their uncertainties, except for $\delta t = 1~\mu s$.

\subsection{Measurement of the impact of the finite resolution of the DDS }\label{intro}

\begin{figure}[h!]
 \centering
 \includegraphics[width=9cm]{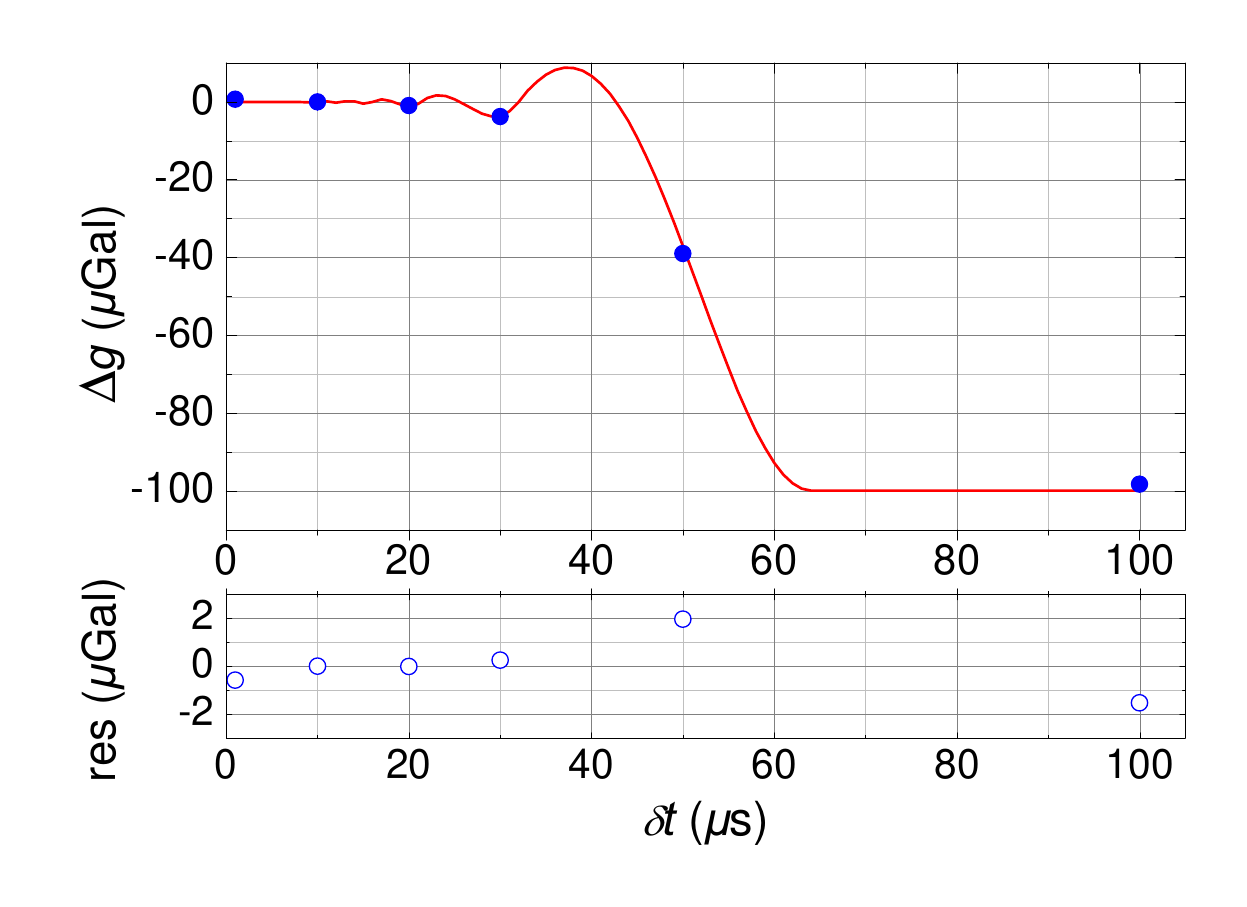}
 \caption{Measurement of $\Delta g$ as a function of $\delta t$ the time step of the chirp (in blue). The continuous line in red is the calculated bias, which is described in the previous section. The residues between theory and measurements are displayed at the bottom as blue open circles.}
 \label{mesdg}
\end{figure}

More interesting, we performed differential $g$ measurements (as in equation \eqref{eqDeltag}) as a function of $\delta t$, using $\delta t_{ref} = 10~\mu$s as a reference parameter, in order to probe the influence of the finite duration of the time step of the chirp. The results are displayed on figure \ref{mesdg}. They show a good agreement with the results of the calculation, which are represented in red. The residues, also plotted on figure \ref{mesdg}, are lower than $2~\mu$Gal even for the largest biases.

\medskip

As discussed in section \ref{Calcul} we then studied the possibility to modulate this effect by either modifying the offset $\Delta t$ or by changing the duration of the Raman pulses $\tau$.  

\begin{figure}[h!]
   \begin{minipage}[c]{.46\linewidth}
      \includegraphics[width=8.5 cm]{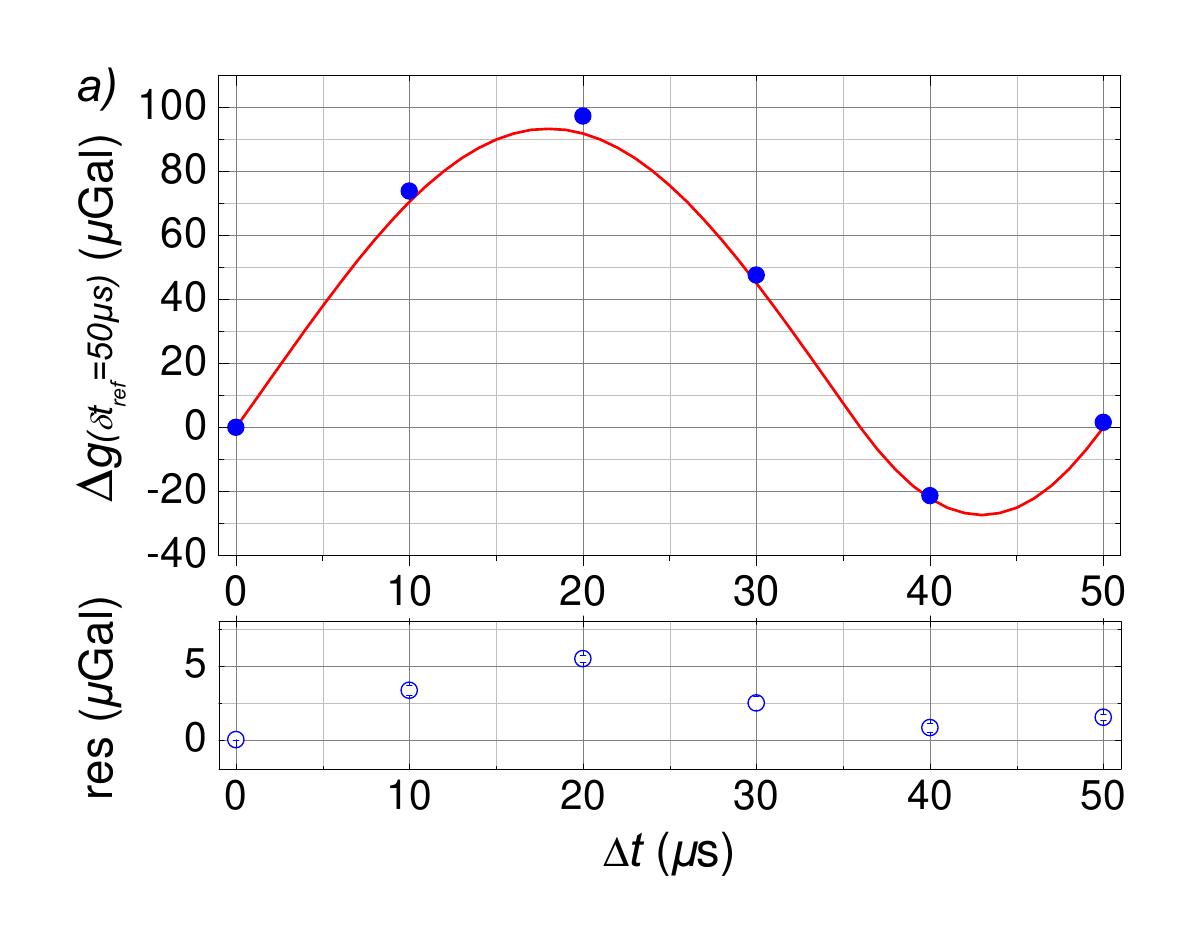}
   \end{minipage} \hfill
   \begin{minipage}[c]{.46\linewidth}
      \includegraphics[width=8.5 cm]{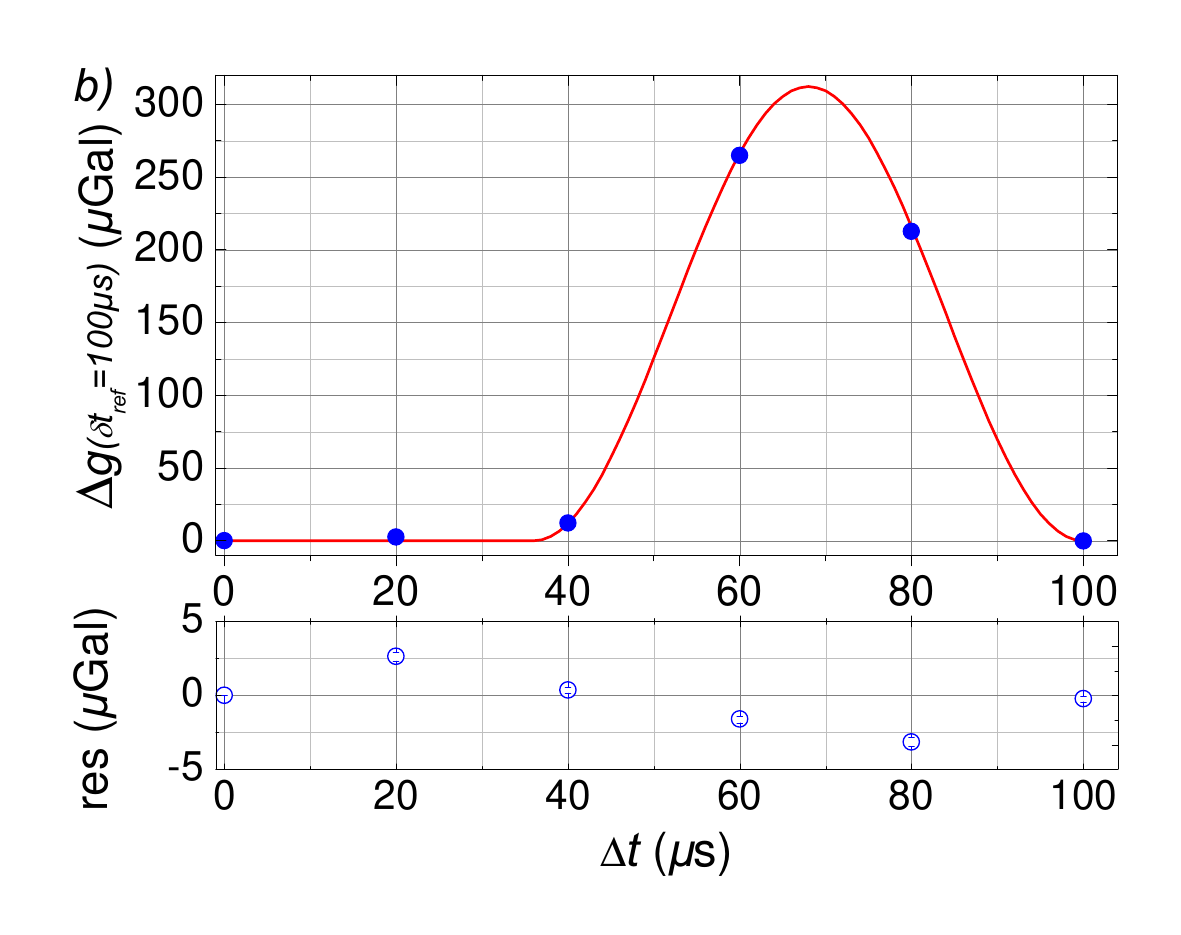}
   \end{minipage}
   \caption{Measurement of $\Delta g$ as a function of the offset $\Delta t$ between the starts of the chirp and the interferometer (in blue circles). a) the reference is $\delta t =50~\mu$s and b) $\delta t = 100~\mu$s.} 
   \label{mesOffset}
\end{figure}

We started by varying the offset $\Delta t$, from 0 to $\delta t$, for two different values of $\delta t =50$ and $100~\mu$s. These two examples were chosen because of the large biases they are expected to induce. Indeed, according to the calculations performed above, this bias is expected to oscillate with a maximum amplitude of only $0.5~\mu$Gal for $\delta t =10~\mu$s, which would be hard to resolve. The measurements were performed in a differential manner with $\delta t =50~\mu$s ($100~\mu$s respectively) and $\Delta t_{ref}=0$ as a reference. The results of these measurements are plotted as blue circles on figure \ref{mesOffset} (a) $\delta t =50~\mu s$ and b) $\delta t =100~\mu s$), together with the results of the calculation, displayed as continuous red lines. Once again the measurements agree with the numerical calculations.  

Finally, we tried to cancel this bias for the particular case of $\delta t =50~\mu$s, by using a Raman sequence of $\tau -2\tau-\tau = 25~\mu$s$ - 50~\mu$s$ -25~\mu$s. We use here the classical sequence $\tau -2\tau-\tau = 16~\mu$s$ - 32~\mu$s$ -16~\mu$s and $\delta t= 50~\mu$s as a reference. We obtain a difference of $37.3 (1)~\mu$Gal close to the expected bias of $37.1~\mu$Gal. The measured bias on the difference being comparable to the expected bias obtained with the simulation, we are confident that the bias is indeed null for the (25-50-25) configuration.

\section{Conclusion}

We have evaluated the impact of the finite temporal resolution of the direct digital synthesizer we use to chirp the Raman laser frequency difference during the interferometer in our cold atom gravimeter. This chirp is essential to maintain the resonance condition of the Raman lasers during the free fall of the atoms. We have discussed and measured the truncation error, of $1/2$ bit on average, arising from the finite frequency resolution of the DDS. Increasing the duration of the time step $\delta t$ decreases the amplitude of the truncation, and improves the resolution on the interferometer phase, which could be useful in a very low noise environment.

However, this results in larger frequency deviations with respect to the ideal linear chirp, inducing potentially large biases on the interferometer phase. This effect can in principle be compensated for by a proper choice of either the duration of the Raman pulses or the time delay between the chirp sequence and the interferometer sequence. Our typical measurement conditions $\delta t= 10~\mu$s results from a compromise between the resolution and the bias on the interferometer phase. The truncation error, of $4.15~\mu$Gal, is rejected by the $k$-reversal algorithm, whereas the bias arising from the frequency error is as small as $0.06~\mu$Gal, well below the current accuracy of the instrument, of order of $2~\mu$Gal \cite{Karcher2018}. But, other choices for the time step can lead to much larger biases, from which $g$ measurements would need to be corrected.

We finally stress that the calculations have been performed with an instrument transfer function which corresponds to square shaped Raman laser pulses. As a follow up study, one could investigate the impact of either more realistic or deliberately shaped pulses \cite{Bess} onto the errors related to the finite resolution of the frequency chirp.

\section{Acknowlegments}
This work has been supported by Region Ile-de-France in the framework of DIM SIRTEQ (CAUCAG project).

\end{document}